\documentclass{ws-procs975x65}

\begin{document}

\title{Covariant Constructive Gravity}
\author{Tobias Reinhart$^*$ and Nils Alex}

\address{Department Physik, Friedrich-Alexander Universit{\"a}t Erlangen-N{\"u}rnberg,\\
91058 Erlangen, Germany\\
$^*$E-mail: tobi.reinhart@fau.de}

\begin{abstract}
We present a method of constructing perturbative equations of motion for the geometric background of any given tensorial field theory.
Requiring invariance of the gravitational dynamics under spacetime diffeomorphisms leads to a PDE system for the gravitational Lagrangian that can be solved by means of a power series ansatz.
Furthermore, in each order we pose conditions on the causality of the gravitational equations, that ensure coevolution of the matter fields and the gravitational background is possible, i.e. gravitational equations and matter equations share the same initial data hypersurfaces.
\end{abstract}


\bodymatter

\section{Introduction}

Given a matter field theory in terms of an action functional $S_{mat}[\Phi;G)$ for a tensorial matter field $\Phi$ whose geometric background is provided by an additional tensorfield $G$, the problem of constructive gravity consists of completing the matter field equations with EOM for $G$ that make the coupled system of matter dynamics and supplemented gravitational dynamics a predictive theory.
In contrast to the existing canonical framework described in Ref.~\refcite{cgg18} , we outline in the following how this problem can be tackled perturbatively from the spacetime covariant point of view.

\section{Perturbative diffeomorphism invariance}
Let $M$ be a smooth 4-dimensional manifold that represents spacetime. The gravitational field is described as a tensorfield $G$ on $M$, i.e. as a section of a bundle $\pi_F : F \rightarrow M$, with adapted coordinate functions $(x^m, v_A)$, where $F \subset T^m_nM$ for some $m$ and $n$. The equations we want to construct for $G$ shall be given as Euler-Lagrange-Equations of an up to now unknown Lagrangian $L : J^1F \rightarrow \Lambda^4 M$, which is a bundle map from the first order jet bundle over $F$ with coordinates $(x^m,v_A,v_{Ap})$ to the bundle of volume forms on $M$.
One of the key requirements we pose on the to-be-constructed gravitational EOM is the invariance under spacetime diffeomorphisms.
The action of $\text{Diff}(M)$ on M extends naturally to  $F$ by pullback and can then be lifted to the first jet bundle in standard ways.\cite{gm04} Doing so, given $f \in \text{Diff}(M)$ that acts on $M$, we can construct bundle automorphisms $f_{\ast}$ and $j^1f_{\ast}$ that act on $F$ and $J^1F$, respectively. Diffeomorphism invariance of the theory is given if the Lagrangian is equivariant w.r.t.\ the lifted action on $J^1F$ and the action by pullback on $\Lambda^4M$, i.e. for all $f \in \text{Diff}(M)$ it holds that:
\begin{align} \label{equi}
    L \circ j^1f_{\ast} = f_{\ast} \circ L .
\end{align}
On the Lie algebra level, (\ref{equi}) can be shown to be equivalent to the following set of first order partial differential equations for $L$:\ \cite{gm04} 
\begin{equation} \label{dens}
\begin{aligned} 
   0 &= L_{:m} \\
   0 &= L^{:A} C^{Bm}_{An} v_B + L^{:Ap} C^{Bm}_{An} v_{Bp} - L^{:Am} v_{An} + L \delta^m_n\\
   0 &= L^{:A (p \vert } C ^{B \vert m)}_{An} v_B,
\end{aligned}
\end{equation}
where the constant tensor $C^{Bm}_{An}$ is obtained from the lie-algebra action of vector fields $\xi$ on $F$, $\xi_F = \xi^m \partial_m + C^{Bm}_{An}V_B \partial_m \xi ^n \partial^{A}$ and hence depends on the specific nature of $G$.
Here and in the following we use the notation $L_{:m} := \partial_m L$, $L^{:A} := \partial^A L$, etc.
Later on we are going to use the equivariance condition of the Lagrangian to explicitly construct gravitational EOM. It is then often advantageous to work directly on the level of the EOM.
Given a Lagrangian $L(x^m,v_A,v_{Ap})$ we obtain the corresponding EOM by applying the variational derivative to $L$ and equating to zero: 
\begin{align}
\begin{aligned}
    0 = &E^A(x^m,v_A,v_{Ap},v_{Apq}) := \frac{\delta L}{\delta v_A} = L^{:A} - D_m(L^{:Am}) \\
    &D_m := \partial_m + v_{Am} \cdot \partial^A + v_{Apm} \cdot \partial^{Ap}. 
\end{aligned}
\end{align}
The EOM are now described by a function $E^A$ on $J^2F$.  
As consequence of ($\ref{equi}$) the EOM satisfy the following first order partial differential equation system:
\begin{equation} \label{tens}
\begin{aligned} 
    0 = &E^A_{:m} \\
    0 = &E^{A:B} C^{Cm}_{Bn} v_C + E^{A:Bp} C^{Cm}_{Bn} v_{Cp} - E^{A:Bm} v_{Bn} \\ 
        &+ E^{A:Bpq} C^{Cm}_{Bn} v_{Cpq} - 2E^{A:Bpm} v_{Bpn} +  E^A \delta^m_n + E^B C^{Am}_{Bn}\\
    0 = &E^{A:B (p \vert } C ^{C \vert m)}_{Bn} v_C + E^{A:B(p \vert q} C^{C \vert m)}_{Bn} v_{Cq} - E^{A:Bpm} v_{Bn}\\
    0 = &E^{A:B (pq\vert } C^{C \vert m)}_{Bn} v_C  .
\end{aligned}
\end{equation}
The problem of constructing diffeomorphism invariant theories for a given tensor field $G$ is thereby translated into the equivalent problem of finding solutions to the equivariance equations (\ref{dens}) on the Level of the Lagrangian or the equations (\ref{tens}) on the level of the EOM. Details concerning the derivation of these equations can be found in Ref. ~\refcite{covcgg19}.
In most cases of physical interest, obtaining general solutions to either one of the two systems of equations is strikingly difficult. Therefore we are going to construct perturbative solutions, more precisely we are seeking power series expansions of the Lagrangian (or the EOM) that solve the equations (\ref{dens}) (or the equations (\ref{tens})) up to the desired order. \cite{seiler95} 
We choose an expansion point $x_0$ that is induced by the flat Minkowski metric $\eta_{ab}$, i.e. $x_0=(x^m,N_A,0,..)$, where the coordinates $N_A = N_A(\eta_{ab})$ are functions of the flat Minkowski metric and the derivative coordinates are zero. This allows the perturbatively constructed theory of gravity to be interpreted as an expansion around flat Minkowski spacetime.
Discharging any explicit $x^m$ dependency that is prohibited due to the first equation in (\ref{dens}) from the very beginning, the most general expansion of the Lagrangian is given as:
\begin{equation} \label{expansion}
\begin{aligned}
    L &= L^{:A: B} \big \vert _{x_0} (v_A-N_A)(v_B-N_B) + L^{:Ap:Bq} \big \vert _{x_0}v_{Ap}v_{Bq} \\
      & \hphantom{=} + L^{:A:B:C} \big \vert _{x_0} (v_A-N_A)(v_B-N_B)(v_C-N_C) \\
      & \hphantom{=} + L^{:Ap:Bq:C} \big \vert _{x_0}v_{Ap}v_{Bq}(v_C-N_C) + \mathcal{O}(4).
    \end{aligned}
\end{equation}
Here any linear terms are neglected as they do not contribute to the EOM. Terms with an odd number of indices are neglected as these can otherwise be eliminated using a result that can be derived from the equivariance equations: \cite{covcgg19} If the chosen expansion point is eta-induced, the expansion coefficients in (\ref{expansion}), i.e. $L^{:A:B}\vert_{x_0}$, etc. are linear combinations of sums of products of $\eta^{ab}$ and $\epsilon^{abcd}$. 
Inserting the induced expansion of the EOM into (\ref{tens}) and evaluating at $x_0$ yields a set of linear equations for the expansion coefficients that contribute to the EOM in linear order. Prolonging (\ref{tens}), i.e. deriving with respect to the coordinates of $J^2F$, inserting the expansion of the $E^A$ and again evaluating at $x_0$, we obtain a system of linear equations for the second order expansion coefficients. In total, by doing so we end up with the following set of linear equations for the linear order expansion coefficients:
\begin{equation} \label{Lin}
    \begin{aligned}
    0 &= L^{:A:B} \big \vert _{x_0} C^{Cm}_{Bn} N_C \\
    0 &= \big \vert_{(pqm)} L^{:Ap:Bq} \big \vert _{x_0} C^{Cm}_{Bn} N_C ,
    \end{aligned}
\end{equation}
where $... = \big  \vert_{(pqm)}$ denotes the total symmetrization in $pqm$ for the whole equation.
In quadratic order we obtain:
\begin{equation} \label{Quad}
    \begin{aligned}
    0 = &-2L^{:A:B}\big \vert _{x_0} C^{Cm}_{Bn} + 6 L^{:A:B:C}\big \vert _{x_0} C^{Dm}_{Bn}N_D + 2 L^{:B:C}\big \vert _{x_0}C^{Am}_{Bn} - 2L^{:A:C}\big \vert _{x_0}\delta^m_n \\
    0 = &\big\vert_{(pq)} 2L^{:Ap:Cm}\big \vert _{x_0}\delta^q_n + L^{:Ap:Bq}\big \vert _{x_0} C^{Cm}_{Bn} + L^{:Ap:Cq:B}\big \vert _{x_0} C^{Dm}_{Bn} N_D \\
        &+L^{:Bp:Cq}\big \vert _{x_0} C^{Am}_{Bn} + L^{:Ap:Cq}\big \vert _{x_0} \delta^m_n \\
    0 = &\big \vert_{(pm)} L^{:Ap:Cm}\big \vert _{x_0} \delta^q_n -2 L^{:Ap:Bq}\big \vert _{x_0} C^{Cm}_{Bn} \\
        &+ \bigl[ L^{:Bp:Cq:A}\big \vert _{x_0} - L^{:Ap:Cq:B}\big \vert _{x_0} - L^{:Bp:Aq:C}\big \vert _{x_0}  \bigr] C^{Dm}_{Bn} N_D\\
    0 = &\big \vert_{(pqm)} L^{:Ap:Bq}\big \vert _{x_0} C^{Cm}_{Bn} + L^{:Ap:Bq:C}\big \vert _{x_0} C^{Dm}_{Bn}N_D.
    \end{aligned}
\end{equation}
Note that the complex quest of solving the PDE system (\ref{tens}) which guarantees diffeomorphism invariance is now translated, at least perturbatively, into the much simpler task of solving a system of linear equations. Furthermore the presented equations take this form irrespective of the specific gravitational field $G$. The only quantity that depends on $G$ is the constant tensor $C^{Am}_{Bn}$ that can easily be computed once $G$ is specified. 
Solving this linear system can be simplified even further by again exploiting the fact that the expansion coefficients are built from $\eta^{ab}$ and $\epsilon^{abcd}$. We construct the most general such expression for each coefficient. These expressions obviously involve undetermined constants. Inserting into (\ref{Quad}) yields a system of linear equations for these constants.
\section{Examples of the construction procedure}
In the following we present two examples of the previously described construction procedure.
\subsection{Metric gravity}
Given the action functional of a Klein-Gordon matter field on a fixed metric background
\begin{align}
S_{\text{KG}}[\Phi;g_{ab})=\int\mathrm{d}^4x\sqrt{-g}g^{ab}\partial_a \Phi \partial_b \Phi,
\end{align}
we want to construct a second order expansion of diffeomorphism invariant EOM for $g_{ab}$ around $\eta_{ab}$. 
Starting with the linear order equations (\ref{Lin}), the first step is computing the most general expressions built from $\eta^{ab}$ and $\epsilon^{abcd}$ for the two linear expansion coefficients. Using the induced symmetries of these one readily finds that: 
\begin{align}
            L^{:ab:cd}\big \vert_{x_0} = & \mu_1 \cdot  \eta^{ab}\eta^{cd} + \mu_2 \cdot \eta^{c (a} \eta^{b) d}.
\end{align}
In similar ways we compute the expression for $L^{:abp:cdq}$ with constants $\nu_1,..,\nu_4$.
Inserting in (\ref{Lin}) and solving for the unknown constants we get:
\begin{align}
    \begin{aligned}
    &\mu_1 = 0 , \  \mu_2 = 0\\
    &\nu_2 = -\nu_1, \ \nu_3 = 2\nu_1 \ \nu_4 = -2\nu_1.
    \end{aligned}
\end{align}
Performing the same computation for the quadratic order expansion coefficients and the corresponding equations (\ref{Quad}), we are left with one overall constant $\nu_1$ which is irrelevant once the EOM are equated to zero. Furthermore the two terms in the expansion of $E^A$ that have no contributions of derivatives of $g_{ab}$ are identically zero.\footnote{In the following such terms are referred to as mass terms and the corresponding constants are called masses, whereas the remaining terms and constants are called kinetic terms and gravitational constants respectively.}
Comparing the obtained quadratic order expansion of the EOM with the Einstein field equations we conclude that the two systems coincide.
\subsection{Area metric gravity}
As a second example we consider the action functional of general linear electrodynamics, in short GLED. The action is given as 
\begin{align}
S_{\text{GLED}}[A_m;G_{abcd}) = \int \mathrm{d}^4x \omega_G G^{abcd} F_{ab} F_{cd},
\end{align}
where $A$ is the electromagnetic potential, $F=dA$ is the field strength, $G$ is the so called area metric, a (0,4) tensor field with the symmetries $G_{abcd} = G_{cdab} = -G_{bacd}$ and $\omega_G$ is a density of weight 1 constructed from $G$. The components of the inverse area metric $G^{abcd}$ are defined by $G^{abcd}G_{cdef} = 4 \delta^{[a}_e \delta^{b]}_f$.
We choose $N_{abcd} = \eta_{ac}\eta_{bd} - \eta_{ad}\eta_{bc} - \epsilon_{abcd}$ as expansion point. With this choice, we recover---to zeroth order in the gravitational field---Maxwell electrodynamics on a Minkowski background.
As before we construct ans{\"a}tze for the expansion coefficients from $\eta^{ab}$ and $\epsilon^{abcd}$. Solving the equations (\ref{Lin}) and (\ref{Quad}) we obtain 3 masses and 7 gravitational constants in linear order, 5 additional masses and 36 additional gravitational constants in quadratic order. Details regarding this particular example can be found in Ref.~\refcite{covcgg19}.
\section{Causal compatibility of gravitational and matter field equations}
Apart from diffeomorphism invariance, the second crucial requirement that constructive gravity poses on the gravitational dynamics is their causal compatibility with the given matter theory.\cite{cgg18} Physically meaningful matter theories have hyperbolic EOM, i.e. given initial data on an initial data hypersurface in $M$, the EOM can be used to evolve this initial data and thereby uniquely predict the values of the matter field away from the hypersurface. 
Assuming this for the given matter theory, one can construct a certain polynomial function on the symmetric product $S^k(T^{\ast}M)$, the so called principal polynomial of the matter equations  $P$. From the vanishing set of $P$ a convex cone in each $T^{\ast}_xM$ is constructed. It can be shown that precisely those hypersurfaces that have at each point a co-normal vector lying inside this cone are admissible initial data hypersurfaces.\cite{cgg18,seiler95} If the gravitational EOM shall complete the given matter dynamics to a predictive theory, we must require that there exist initial data hypersurfaces that are common to both EOMs. This is certainly the case if the principal polynomial of the gravitational EOM has the same vanishing set as the principal polynomial of the matter theory. In the following we are going to explore how this additional requirement might or might not restrict the gravitational constants and masses that were obtained in the previous two examples. We start with metric gravity. The principal polynomial of the Klein-Gordon theory is, up to an overall factor, simply given by 
\begin{align}
P_{\text{KG}}(k) = g^{ab}k_ak_b = \eta^{ab}k_a k_b - \eta^{ac}\eta^{bd}h_{cd} k_a k_b  + \mathcal{O}(2),
\end{align}
where we introduced $h_{ab} := g_{ab}-\eta_{ab}$ and expanded the polynomial in powers of $h_{ab}$. Computing the principal polynomial of the perturbative expansion of metric gravity obtained before, we find that
\begin{align}
    P_{\text{metric}}(k) = \bigl \{ (\eta^{ab} k_a k_b - \eta^{ac}\eta^{bd}h_{cd} k_a k_b) \bigr \} ^2  + \mathcal{O}(2) .
\end{align}
Obviously this polynomial describes the same vanishing set as $P_{\text{KG}}$. Hence we get no additional condition from the causality requirement. This is not too surprising as except from an overall gravitational constant there is no freedom left in the perturbative metric EOM in the first place. \\
The principal polynomial of GLED was first computed by Rubilar \cite{rub02} and is given by 
\begin{align} \label{GLEDPoly}
\begin{aligned}
    P_{\text{GLED}}(k) &= -\frac{1}{24}\omega_G^2\epsilon_{mnpq}\epsilon_{rstu}G^{mnra}G^{bpsc}G^{dqtu}k_ak_bk_ck_d \\
                &= \bigl[ 1 -  A \eta(H)- \frac{1}{2} A \epsilon(H) + \frac{1}{12} \epsilon(H) \bigr] \eta(k)^2 - H(k)\eta(k) + \mathcal{O}(2)\\
                &= \bigl\{  \bigl[ 1 - \frac{1}{2} A \eta(A) - \frac{1}{4} A \epsilon(H) +  \frac{1}{24} \epsilon(H) \bigr] \eta(k) - \frac{1}{2} H(k)       \bigr\}^2 + \mathcal{O}(2),
\end{aligned}
\end{align}
where $H_{abcd}:=G_{abcd}-N_{abcd}$, $\eta(H) := \eta^{ac}\eta^{bd}H_{abcd}$, $\epsilon(H):=\epsilon^{abcd}H_{abcd}$, $H(k):=\eta^{ac}\eta^{bp}\eta^{cq}H_{abcd}k_pk_q$, $\eta(k):=\eta^{pq}k_pk_q$ and $A$ is a constant that depends on the specific choice for the density $\omega_G$. Note that, as we are only interested in the vanishing set of the principal polynomial, we can multiply a given polynomial with a non vanishing density of arbitrary weight. Such a density admits the general form
\begin{align}
X = 1+ b_1 \cdot (\eta(H) + \frac{1}{2} \epsilon(H)) + \frac{b_2}{12}\epsilon(H) + \mathcal{O}(2),
\end{align}
for arbitrary constants $b_1$ and $b_2$.
Multiplying  (\ref{GLEDPoly}) with such a density we see that perturbatively the GLED polynomial describes the same vanishing set as 
\begin{align} \label{GLEDPoly2}
\begin{aligned}
    \widetilde P_{\text{GLED}}(k) = X \cdot P_{\text{GLED}} = 
    \bigl\{  \bigl[ 1 - \frac{1}{2} (A-b_1) \eta(H) - \frac{1}{4} (A-b_1) \epsilon(H)) + \\ \frac{1+b_2}{24} \epsilon(H) \bigr] \eta(k)
    -\frac{1}{2} H(k)       \bigr\}^2 + \mathcal{O}(2).
\end{aligned}
\end{align}
The principal polynomial of the constructed, diffeomorphism invariant expansion of area metric gravity to second order is calculated as
\begin{align} \label{AreaPoly}
    P_{\text{area}} = \bigl\{  \bigl[ 1 - \frac{1}{2} C \eta(H) - \frac{1}{4} C \epsilon(H) +  \frac{7}{12\cdot13} \epsilon(H) \bigr] \eta(k) - \frac{1}{2} H(k)       \bigr\}^{13} + \mathcal{O}(2),
\end{align}
where $C$ is an expression that depends on the gravitational constants of the theory, but not on $H_{abcd}$. \cite{covcgg19} 
Comparing (\ref{GLEDPoly2}) and (\ref{AreaPoly}) shows that, multiplying the GLED polynomial with a density with constants $b_1 = A -C$ and $b_2 = \frac{1}{13}$, the two polynomials are products of the same factor and therefore describe the same vanishing set in $\mathcal{O}(2)$. Hence already the required diffeomorphism invariance fixes the principal polynomial of area metric gravity to the point where, perturbatively, the EOM have the same initial data hypersurfaces as those of GLED.

\end{document}